# Reduced crystal symmetry as origin of the ferroelectric polarization within the incommensurate magnetic phase of TbMn$_2$O$_5$

N. Narayanan[1,2], P. J. Graham[3], P. Rovillain[2,3,4], J. O'Brien[3], J. Bertinshaw[3,5], S. Yick[2,3], J. Hester[2], A. Maljuk[6,7], D. Souptel[7], B. Büchner[7,8,9], D. Argyriou[6,10], and C. Ulrich[3,*]

[1]*School of Physical, Environmental and Mathematical Sciences, University of New South Wales, Canberra, Australian Capital Territory 2600, Australia*

[2]*The Australian Centre for Neutron Scattering, Australian Nuclear Science and Technology Organisation, Lucas Heights, New South Wales 2234, Australia*

[3]*School of Physics, University of New South Wales, Sydney, NSW 2052, Australia*

[4]*Sorbonne Universitée, CNRS, Institut des NanoSciences de Paris, INSP, UMR7588, F-75005 Paris, France.*

[5]*Max-Planck-Institut für Festkörperforschung, Heisenbergstrasse 1, D-70569 Stuttgart, Germany*

[6]*Helmholtz-Zentrum Berlin für Materialien und Energie, D-14109 Berlin, Germany*

[7]*Leibniz Institut für Festkörper- und Werkstoffforshung (IFW) Dresden, Helmholtzstr. 20, D-01069 Dresden, Germany*

[8]*Institut für Festkörper- und Materialphysik, Technische Universität Dresden, 01069 Dresden, Germany*

[9]*Würzburg-Dresden Cluster of Excellence ct.qmat, Technische Universität Dresden, 01062 Dresden, Germany*

[10]*European Spallation Source ESS AB, S-22100 Lund, Sweden*

(Dated: 11 September 2021)

**ABSTRACT**

The precise crystal symmetry and hence the emergence of the electric polarization still remains an open question in the multiferroic materials $R$Mn$_2$O$_5$ ($R$ = rare-earth, Bi, Y). While previous diffraction studies have indicated that $R$Mn$_2$O$_5$ possesses the centro-symmetric space group P*bam*, an atomic displacement allowing for the electric polarization would require a non-centrosymmetric crystal symmetry. Our single crystal neutron diffraction experiments on TbMn$_2$O$_5$ provide direct evidence of a reduced crystallographic symmetry already above the magnetic and ferroelectric phase transitions and a change in magnetic order upon entering the ferroelectric phase. This is indicated through the presence of additional nuclear Bragg reflections that are otherwise forbidden for the space group P*bam* but are in good agreement with the polar space group P*12$_1$1*. It implies that the exchange-striction, which arises from a symmetric $\mathbf{S_i} \cdot \mathbf{S_j}$ spin coupling, is the dominating mechanism for the generation of the electric polarization in the commensurate magnetic phase of TbMn$_2$O$_5$. Furthermore, the commensurate magnetic reflections are in accordance with a quartile step spin-spiral along the *c*-axis. Therefore, the antisymmetric $\mathbf{S_i} \times \mathbf{S_j}$ exchange via the inverse Dzyaloshinskii-Moriya interaction contributes as well and becomes the leading term in the low temperature incommensurate spin-spiral magnetic phase. These new findings provide important information for the understanding of the complex interplay between the magnetic and the structural order throughout the $R$Mn$_2$O$_5$ series of type-II multiferroics.

**I. INTRODUCTION**

The coexistence of both ferroelectricity and magnetism in multiferroic materials offers new possibilities in developing novel data storage and sensor technologies with advanced functionalities since both ordered components can be exploited simultaneously [1-5]. In type-II multiferroics, both properties can even be switched by each other, known as the magnetoelectric (ME) effect. The direct control of magnetic components in memory storage devices through electric manipulation can provide faster storage and retrieval of information and has the potential to increase the data storage capacity by 5-6 orders of magnitude [3]. However, due to the complex spin and charge interactions in these systems, the underlying quantum mechanical processes behind these phenomena are still not fully understood.

Although there is a general understanding of this phenomenon, some unsettled discussion remains concerning the role of ionic displacements and the lattice in general, in influencing the ME coupling in a number of multiferroic systems [4,5]. A more thorough understanding into the underlying conditions for ME coupling will thus provide valuable insight into multiferroic systems and further the progress in their applications.

The $R$Mn$_2$O$_5$ series ($R$ = rare-earth Pr - Lu, and Bi, Y) is a group of such multiferroic systems demonstrating a colossal ME effect that has so far attracted considerable efforts in scientific investigation [6-11]. While the electric polarization of around 45 nC/cm$^2$ is several orders of magnitude weaker as compared to regular ferroelectrics, its ME coupling is strong [12] and the direction of the electric polarization can even be reverted in a magnetic field applied along the crystallographic $a$-direction [7]. Already in the 1960s Bertaut *et al.* [13,14] and Abrahams *et al.* [15] determined that the space group of $R$Mn$_2$O$_5$, with $R$ = Ho, Bi, and Dy respectively, is P*bam* (#55 in the 'International Tables for Crystallography' [16]). This space group was confirmed for the entire series of $R$Mn$_2$O$_5$ [17-23] (for review see [24,25]). The unit cell consists of four formula units and the lattice parameters are $a$ = 7.3251(2), $b$ = 8.5168(2), and $c$ = 5.6750(2) at room temperature [20]. The corresponding crystallographic structure is shown in Fig. 1 and can be described as a network of edge and corner sharing Mn$^{4+}$O$_6$ octahedra (Mn$^{4+}$: 3d$^3$; S=3/2 at the Wyckoff position 4h) and base centred Mn$^{3+}$O$_5$ square pyramids (Mn$^{3+}$: 3d$^4$; S=2 at the Wyckoff position 4f). Along the $c$-axis the Mn$^{4+}$O$_6$ octahedra form edge-sharing chains which are linked by Mn$^{3+}$O$_5$ pyramid pairs. The $ab$-plane is dominated by Mn$^{3+}$O$_5$ - Mn$^{4+}$O$_6$ - Mn$^{3+}$O$_5$ corner-sharing (Mn$^{3+}$-Mn$^{4+}$-Mn$^{3+}$) units, which form zigzag chains in $a$-direction (see light green lines in Fig. 1), and which are weakly connected along the $b$-direction. From another perspective, five Mn$^{3+/4+}$ polyhedral (two octahedra and three pyramids) are forming a circular arrangement in the $ab$-plane with the sequence Mn$^{4+}$-Mn$^{3+}$-Mn$^{3+}$-Mn$^{4+}$-Mn$^{3+}$ (see red circle in Fig. 1). Since all nearest neighbour interactions are AFM, this five-spin structure can be considered as the origin of the geometrical spin frustration in the compounds $R$Mn$_2$O$_5$.

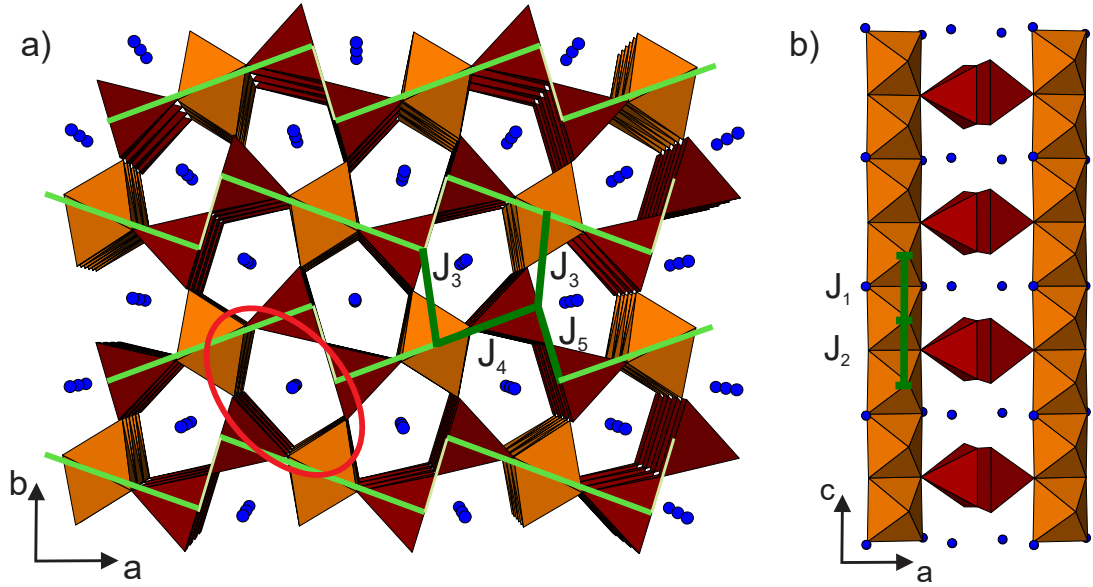

*Fig 1: Crystal structure of TbMn$_2$O$_5$ at room temperature (crystallographic data from Ref. [20]). (a) ab-plane. The key elements are edge-sharing Mn$^{3+}$O$_5$ - Mn$^{4+}$O$_6$ - Mn$^{3+}$O$_5$ pyramid-octahedron-pyramid units which are linked to zigzag chains along the a-direction (see light green lines). The red circle indicates the circular arrangement of five Mn$^{4+}$-Mn$^{3+}$-Mn$^{3+}$-Mn$^{4+}$-Mn$^{3+}$ spins, which are the origin of the spin frustration in TbMn$_2$O$_5$. (b) ac-plane. Edge sharing Mn$^{4+}$O$_6$ octahedral chains along the c-direction are linked by two Mn$^{3+}$O$_5$ octahedra. The dark green lines in both panels indicate the magnetic exchange interactions J$_1$-J$_5$ in the ab- and ac-planes.*

Five antiferromagnetic superexchange interactions J$_1$-J$_5$ are responsible for the magnetic order in the series $R$Mn$_2$O$_5$ (see dark green lines in Fig. 1). These interactions were already identified by Abrahams *et al.* in 1967 [15] and are labelled according to the notation given in literature [11,15,17,22,23]. The J$_1$ and J$_2$ exchange interactions couple the spins of the Mn$^{4+}$ ions within the one-dimensional chain along the c-axis; J$_1$ is at the level of the Tb$^{3+}$ layers and J$_2$ is connected through the Mn$^{3+}$O$_5$ pyramidal layers. As a consequence, the J$_2$ magnetic interaction is competing with J$_3$ and J$_4$ which leads to the incommensurate magnetic order along the *c*-axis. The exchange interaction J$_4$ acts within the Mn$^{3+}$-Mn$^{4+}$-Mn$^{3+}$ units of the zig-zag chains in the *ab*-plane. J$_3$ connects the zig-zag chains in *b*-direction by a Mn$^{4+}$-Mn$^{3+}$ superexchange interaction through pyramidal basal corners, while J$_5$ connects the Mn$^{3+}$-Mn$^{4+}$-Mn$^{3+}$ units antiferromagnetically in *a*-direction through the pyramidal basal-basal Mn$^{3+}$-Mn$^{3+}$ ions. In the case of the centrosymmetric crystal structure P*bam*, J$_3$ is degenerate. An atomic displacement of the $Mn^{3+}$ or $Mn^{4+}$ ion in the *ab*-plane would lift this degeneracy and lead to two different $J_3$ interactions. According to theoretical first-principles calculations [24,25] and inelastic neutron scattering experiments on $YMn_2O_5$ and $TbMn_2O_5$ [26,27], $J_4$ and $J_5$ are the leading contributions. These interactions are responsible for the antiferromagnetic arrangement of the spins in the zig-zag chains, which are weakly coupled through $J_3$ in the *b*-direction. In the case of a magnetic rare-earth ion, additional exchange interactions between the $R^{3+}$-ions and between the $R^{3+}$ and $Mn^{3+}$ and $Mn^{4+}$-ions are present. However, since the phase diagram of the $RMn_2O_5$ series shows a universal behavior independent of the magnetic nature of the $R^{3+}$-ions [10,11], the $R^{3+}$-ions have only a minor influence on the magnetic structure of the Mn-sublattice and would only become relevant at low temperatures.

Due to the different competing magnetic interactions caused by the geometrical frustration, $TbMn_2O_5$ possesses several magnetic phases with both commensurate (CM) and incommensurate (ICM) spin arrangements [17,22,23,28]. The onset of the antiferromagnetic order occurs at $T_{HT-ICM}$ = 42 K to the high temperature incommensurate magnetic phase (HT-ICM), which is characterized by the magnetic propagation vector $q_{HT-ICM}$ = (1/2-$\delta_x$, 0, 1/4+$\delta_z$). The values of $\delta_x$ and $\delta_z$ decrease continuously with temperature, i.e. the magnetic propagation vector $q_{HT-ICM}$ changes from (0.487, 0, 0.276) to (0.496, 0, 0.262) [6,17,19,22,23,28,29]. A commensurate magnetic (CM) structure is formed at $T_{CM}$ = 37.8 K, where the magnetic propagation vector locks-in to the value of $q_{CM}$ = (½, 0, ¼). This phase transition is accompanied by the onset of an electric polarization along the *b*-axis [7,12]. Below $T_{LT-ICM}$ = 22.4 K the magnetic structure becomes incommensurate again (low temperature incommensurate phase, LT-ICM). At this phase transition the electric polarization drops almost to zero and then increases slightly upon further cooling. The components $q_x$ and $q_z$ of the magnetic propagation vector change continuously and reach a stable value of $q_{LT-ICM}$ = (0.488, 0, 0.313) below T ~16 K. Finally, at $T_{Tb}$ = 9 K the $Tb^{3+}$-moments order magnetically and the electric polarization increases strongly.

In the *ab*-plane the spins of $TbMn_2O_5$ are oriented antiferromagnetically. Their direction is predominantly within the *ab*-plane with a weak canting of the spins out-of-plane [22,23,29]. They are pointing almost perfectly along the direction of the $Mn^{3+}$-$Mn^{4+}$-$Mn^{3+}$ polyhedral units of the zigzag chains, i.e. they possess an angle to the *a*-axis of 10° (14°) for the $Mn^{3+}$ spins and 24° (11°) for the $Mn^{4+}$ according to Chapon *et al.* [22] and Blake *et al.* [23] (in brackets). This holds for all magnetic phases, the HT-ICM, CM, and LT-ICM phases. At low temperature the $Tb^{3+}$ moments induce a further rotation of $Mn^{3+}$ and $Mn^{4+}$ spins away from the *a*-axis. The $Tb^{3+}$ spins are almost parallel to the $Mn^{3+}$ spins and antiferromagnetically arranged in a '- + + -' sequence in the *ab*-plane [6,23,29]. Already in 1973, Buisson *et al.* described the spin structure of the rare-earth series $RMn_2O_5$ with R = Lu, Er, Y, Ho, Tb, and Nd as helical [17]. Recent neutron diffraction experiments on $RMn_2O_5$ with R = Y, Tb, Ho, Er, and Bi have indicated that a spin-spiral exists also in the *bc*-plane with a propagation direction along the *c*-axis and a $\pi/2$ phase shift, i.e. a spin quadrature in the CM phase and close to $\pi/2$ phase steps in the high temperature and low temperature ICM phases [29-37].

In contrast to multiferroics such as $TbMnO_3$ and $DyMnO_3$ [38-40], the strongest electric polarization in the $RMn_2O_5$ series appears in the CM phase and not the ICM phases. This seems to be in contradiction to the model of ferroelectricity induced by magnetic chirality [41,42]. The existence of an electric polarization in the commensurate magnetic phase raises the question about the responsible microscopic mechanism. In general, two models have been proposed for the magnetoelectric coupling: (i) ferroelectric polarization through atomic displacements caused by magnetic exchange-striction and (ii) ferroelectricity induced by a spin-spiral. In case of the magnetic exchange-striction the symmetric Heisenberg exchange interaction causes a displacement of the $Mn^{3+}$ or $Mn^{4+}$ ions. Quantitatively the magnitude of the electric polarization is proportional to the scalar product of $\mathbf{S_i} \cdot \mathbf{S_j}$ [6,22,23,43,44]. The non-collinearity is not essential in this case and the stacking order along the *c*-axis is irrelevant. In the spin-spiral model the ferroelectricity is generated through magnetic chirality [41,42]. A spin-spiral can be caused by the spin-orbit driven antisymmetric Dzyaloshinskii-Moria interaction. A spin current, which is induced by the spin-spiral generates an electric polarization $\mathbf{P} \propto \mathbf{r_{ij}} \times (\mathbf{S_i} \times \mathbf{S_j})$, where $\mathbf{r_{ij}}$ is the propagation direction of the spin-spiral [4,41]. It is important to note that in the case of a spin current, atomic displacements are not necessarily required, and the electric polarization can arise entirely from orbital polarizations. In case of the $RMn_2O_5$ series, the propagation direction of the spin-spiral is along the *c*-direction and the spins are rotating in the *bc*-plane. Hence, the cross product between $\mathbf{S_i}$ and $\mathbf{S_j}$ is pointing along the *a*-direction and the induced electric polarization is along the *b*-direction. Since a spin-spiral has been identified in all magnetic phases, both in the CM as well as in the ICM phases, the spin current model most likely contributes in all magnetic phases to the electric polarization. However, this does not explain why the strongest electric polarization occurs in the CM phase. Therefore, in the CM phase, considerable contributions to the electric polarization must come from magnetic exchange-striction. However, due to symmetry considerations, ferroelectricity is formally forbidden in centrosymmetric, and hence non-polar crystals [46] as in the case of the originally determined space group P*bam*. Therefore, Kagomiya *et al.* [6] proposed through group-theoretical considerations that the symmetry is lowered to the non-centrosymmetric, polar space group P*b2$_1$m* (#26). This is achieved through an atomic displacement of either the $Mn^{3+}$ or $Mn^{4+}$ ions along the zig-zag chains in the *ab*-plane. The resulting electric polarizations would cancel out at the atomic level along the *a*-direction, while they would add up along the *b*-direction. Since the expected atomic displacements responsible for the breaking of the symmetry are in the order of 0.005 Å (see e.g. $DyMnO_3$ and $TbMnO_3$ [40,47]), direct evidence using diffraction techniques is elusive.

Various experiments have been performed thus far to provide a proof for the atomic displacements. Anomalies in the thermal expansion indicate the existence of a strong magnetostriction effect [48,49]. In their neutron diffraction experiments Chapon *et al.* [22] observed a significant increase of the thermal displacement parameters of the $Mn^{3+}$ ions (see also Refs. [23,50,51]) while Wilkinson *et al.* [29] observed large thermal displacement parameters for the $Mn^{4+}$ ions. This could be an indication for a displacement of either the $Mn^{3+}$ or $Mn^{4+}$ ions and hence a splitting of their sites into two. This is supported by Mössbauer spectroscopy through the observation of two different hyperfine sextets [52,53]. As consequence the magnetic exchange interaction $J_3$ between the zigzag chains in the *ab*-plane would split into two different interactions (see Fig. 1). This would partly lift the magnetic frustration and would contribute to stabilize the magnetic structure. Further evidence for a change in crystal symmetry was provided by optical spectroscopy where an additional phonon mode was observed in infrared spectroscopy whose shift in phonon frequency and intensity scales with the square of the electrical polarization [54]. A first crystallographic indication for a change in the space group towards a polar group allowing ferroelectricity was provided by Noda *et al.* [32] who performed single crystal x-ray synchrotron diffraction experiments on $RMn_2O_5$ with $R$ = Y and Tb. They observed additional fundamental Bragg peaks for (*h* 0 *l*) at h-odd positions, which are forbidden in P*bam* (see Table 1 in the Appendix) and also structural modulations along the *a*- and *c*-axis. Due to the limited number of reflections, the refinement results are ambiguous and they propose either the space group P*b2$_1$m* or P*12$_1$1* for the ferroelectric phase. A more recent paper with evidence for symmetry-breaking was published by Balédent *et al.* [55] who performed single crystal x-ray synchrotron diffraction experiments on $RMn_2O_5$ with $R$ = Pr, Nd, Gd, Tb, and Dy at temperatures above the magnetic and ferroelectric phase transition. They observed additional Bragg peaks for (*h* 0 *l*) and (0 *k* l) at odd positions for *h* and *k*, which are forbidden in P*bam*. In combination with first principles calculations, they proposed the non-centrosymmetric space group P*m*, which allows for a displacement of the $Mn^{4+}$-ions and $O^{2-}$-ions in the *ab*-plane and can generate an electric polarization along the *b*-direction.

However, the mechanism of the magnetoelectric coupling in the series $RMn_2O_5$ is still not fully understood and questions remain about the final crystallographic structure or if a thus far unobserved spin canting, i.e. a non-collinearity exists in the CM phase. Therefore, we have performed high resolution single crystal neutron diffraction experiments on $TbMn_2O_5$ at various temperatures in order to determine the precise crystallographic and magnetic structure in all the magnetic and ferroelectric phases. Compared to powder diffraction, single crystal diffraction experiments offer the advantage that (i) even weakest Bragg reflections peaks can be observed, (ii) each Bragg reflection is independently observed instead of their multiplicity, and (iii) that there is no ambiguity about their indexing. Through the extinction rules our data provide a direct evidence for the symmetry lowered polar space group P*1$2_1$1*. Simultaneously the temperature dependence of the magnetic structure was determined for each magnetic phase.

## II. EXPERIMENTAL DETAILS

Temperature-dependent neutron diffraction experiments were performed on a single-crystal $TbMn_2O_5$. The 4x3x3 $mm^3$ single crystal was grown from $PbO$-$PbF_2$-$B_2O_3$ flux in a Pt crucible similar to the method described in Ref. [56]. The crystal was oriented using the neutron Laue-diffraction instrument JOEY (see Fig. S1 in the Supplemental Material [57]) and then measured on the high flux neutron diffractometer Wombat [58] at the OPAL research reactor at ANSTO. The instrument Wombat utilizes a monolithic $^3$He area detector bank that covers 120° in *2θ* within the equatorial plane and ± 7.8° in µ out of the equatorial plane. A neutron wavelength of λ ≈ 2.41 Å was selected for all experiments using a Ge monochromator at the (113) reflection, which offers the advantage that the second-order reflections are forbidden. However, weak third-order contaminations were observed and identified in the diffraction patterns. Reciprocal space maps in the (*h k* 0), (*h* 0 *l*), and (0 *k l*) planes were obtained with about 60° sample rotations of ω = 0.5° steps at selected temperatures of 4.0 K, 28 K, 35 K, 40 K and 50 K, i.e. in the LT-ICM, CM, HT-ICM, and in the paramagnetic (PM) phases. Additional temperature-dependent scans were performed using a narrower *q*-range with an about 25° sample rotation of ω = 0.25° steps to capture a sufficient number of magnetic and nuclear Bragg reflections in each measurement sequence. Refinements of the integrated intensities were carried out by using a least-square method of the program suite FULLPROF [59].

### III. RESULTS AND DISCUSSION

Figure 2 shows the reciprocal space map (RSM) in the ($h$ 0 $l$)-plane obtained by neutron diffraction using the instrument Wombat at ANSTO. The intense ring structures originate from powder-like diffraction of aluminium from the sample holder and the cryostat. The sharp peaks arise from nuclear and magnetic Bragg reflections of $TbMn_2O_5$ for the different magnetic phases, i.e. the LT-ICM phase at 4 K, the CM phase at 28 K, the HT-ICM phase at 40 K and the paramagnetic phase at 50 K. The nuclear Bragg reflections corresponding to the P*bam* space group are indicated by white right angles and the magnetic Bragg peaks by red right angles. Allowed nuclear reflections in the space group P*bam* appear at positions with $h=2n$ and $l=n$ where $n$ is an integer (see Table T1 in the Supplemental Material [57]). We observe additional weak reflections at ($h$ 0 $l$) positions with odd $h$-indices ($h=2n+1$) and $l$ being an integer and at ($h$ 0 $l$) positions with integer $h$ and half integer $l$-indices ($l=0.5n$). The additional Bragg reflections cannot be attributed to $\lambda/2$ contaminations since a (113) oriented Ge monochromator was used to select the incident wavelength. However, $\lambda/3$ contaminations are possible. Weak reflections are visible for non-integer $h$ and $l$-values at (0 $k$ $l$) positions with $h=1/3$ and $l=1/3$ and are probably caused by $\lambda/3$ contaminations. However, $\lambda/3$ contaminations cannot account for the additional observed Bragg peaks at odd $h$-positions, since for example, the corresponding Bragg peak (-9, 0, 6) which could cause the third-order peak at (-3, 0, 2) is forbidden in the space group P*bam* and would hence be very weak for any symmetry lowered crystal structure. Therefore, the additional Bragg reflections cannot be associated with any higher order contaminations and must be caused by additional nuclear Bragg peaks, which would not be present for the P*bam* crystal structure. This provides strong evidence that the crystal symmetry is lower than P*bam*.

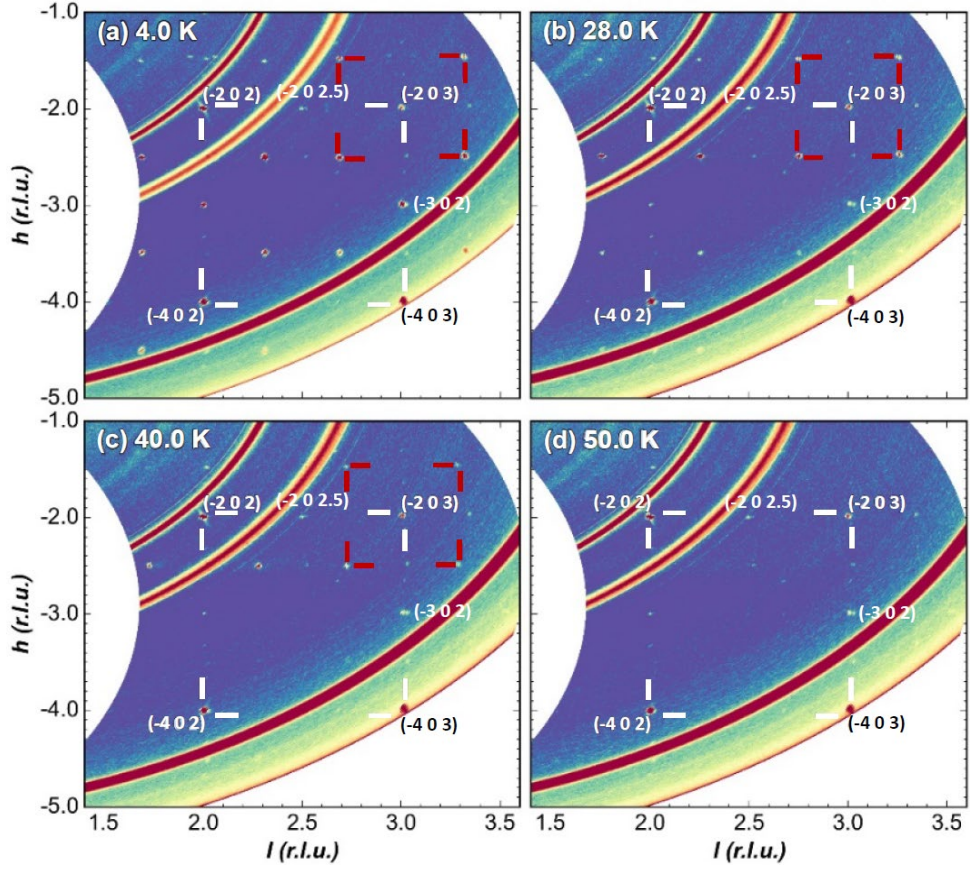

FIG. 2. Reciprocal space maps of TbMn$_2$O$_5$ in the (h 0 l)-plane at various temperatures, i.e. at 4 K in the LT-ICM, at 28 K in the CM, at 40 K in the HT-ICM, and at 50 K in the PM phase. In the Pbam space group nuclear reflections appear at position with h=2n and l=n where n is an integer (white right angles). Additional nuclear Bragg peaks are observed at (h 0 l) positions with h=odd integer and l=integer, and at (h 0 l) positions with h=integer and l=half-integer indices. The magnetic Bragg reflections at (0.5±$\delta_x$, 0, 0.25±$\delta_z$) are indicated by red right angles.

The additional structural Bragg peaks at (*h* 0 *l*) with *h*=odd integer, *l*=integer and *h*=integer, *l*=half-integer indices persist at all temperatures, i.e. throughout the LT-ICM, CM, HT-ICM and PM phases and their intensity remains almost constant at all temperatures. It is important to note that the additional nuclear Bragg reflections are already present at temperatures above the magnetic phase transition, which indicates that the crystal structure persists from room temperature down to the lowest measured temperature. This is in agreement with Balédent *et al.* [55], who also observed additional Bragg reflections at room temperature in their single x-ray synchrotron diffraction experiments on *R*Mn$_2$O$_5$ with *R* = Pr, Nd, Gd Tb, and Dy. In order to obtain a complete overview, the RSMs of the (*h k* 0) and (0 *k l*) planes were measured as well. Figures. 3(a) and 3(b) show the RSM of the *hk*-plane taken at T = 5.5 K and 50 K, i.e. in the LT-ICM and PM phase, respectively. Nuclear Bragg peaks are observed at (*h k* 0) positions with *h* and *k* being an integer. Their intensity does not change with temperature as shown for the (3, -4, 0) reflection in Fig. S2 in the Supplemental Material [57]. These Bragg reflections are in accordance with the P*bam* space group. At the lower temperature additional peaks appear at about half-integer *h*-positions. They disappear above the magnetic phase transition and can therefore be attributed to magnetic scattering. Due to the vertical detector range of ±7.8°, out-of-plane reflections with up to $l = \pm 0.36$ are also detected. Therefore, these Bragg peaks can be assigned to ($h\pm0.51, k, \pm0.31$) magnetic reflections.

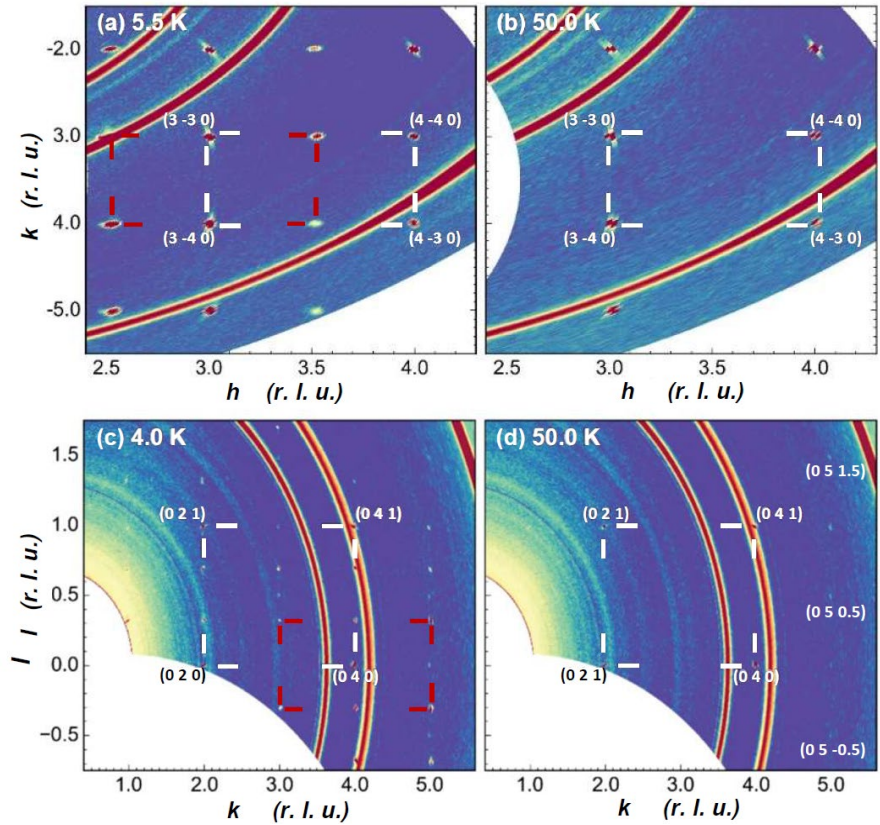

FIG. 3. (a, b) RSM of the hk-plane of TbMn$_2$O$_5$ measured in the LT-ICM phase at 5.5 K and in the PM phase at 50K. The nuclear Bragg reflections are highlighted by white right angles. At low temperature out-of-plane magnetic Bragg peaks at (h±0.51, k, ±0.31) are also visible due to the vertical detector range of ±7.8°. (c, d) RSM of the kl-plane at 4.0 K (LT-ICM phase) and at 50 K (PM phase). Additional nuclear Bragg peaks appear at both temperatures at half-integer l-indices and are labelled underneath the reflection. In the LT-ICM phase at 4 K additional magnetic Bragg peaks appear. They can be attributed to out-of-plane reflections at h = ±0.52 and are indicated by red right angles.

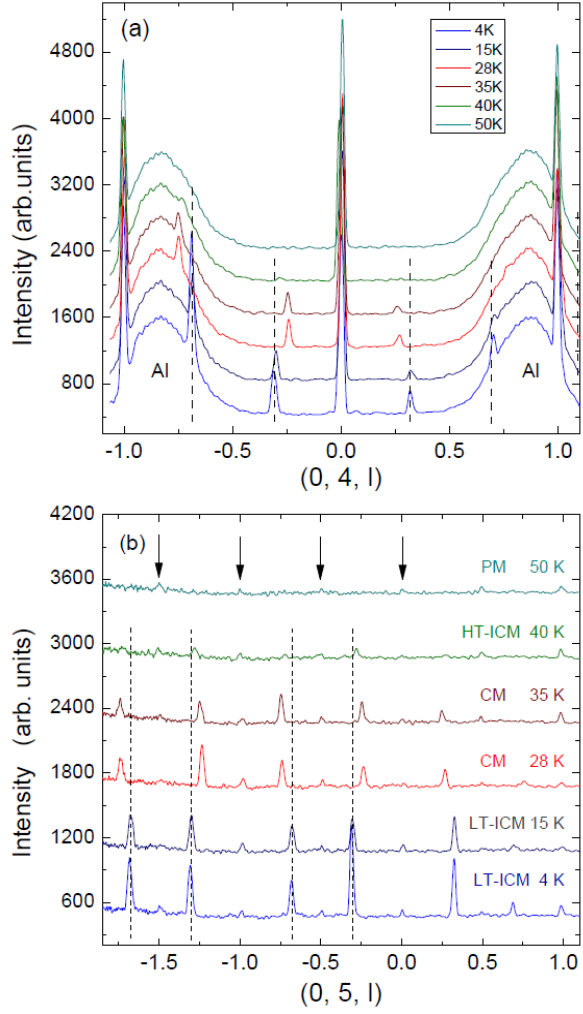

FIG. 4. Line cuts through the RSMs of the kl-plane along the l-direction, (a) along (0 4 l) and (b) along (0 5 l). The out-of-plane magnetic Bragg reflections at (±0.52, k, ¼+$\delta_z$) are indicated by vertical dashed lines. For the (0 5 l) direction, weak structural reflections are observed at l half-integer and l integer positions (see black arrows). They are present at all temperatures, also above the magnetic phase transition and their intensity does not change with temperature.

Figures 3(c) and 3(d) show the RSMs of the *kl*-plane measured at 4 K in the LT-ICM phase and above the magnetic ordering temperature at 50 K. The nuclear Bragg reflections which are in accordance with the space group P*bam* are highlighted by white right angles. Due to the vertical detector range of ±7.8° out-of-plane magnetic Bragg peaks appear at lowest temperature at (±0.52, *k*, ¼+$\delta_z$) in the *kl*-plane (see red right angles). Figure 4 shows line cuts along the *l*-direction extracted from the RSMs of the *kl*-plane for the (0 4 *l*) and (0 5 *l*) directions, respectively. The out-of-plane magnetic Bragg reflections at (±0.52, *k*, ¼+$\delta_z$) are indicated by vertical dashed lines. They show the expected shift in position for the different magnetic phases from $l = 0.278$ in the HT-ICM phase to $l = 0.25$ in the CM phase and finally to $l = 0.31$ in the LT-ICM phase at lowest temperature. Important is that we observe additional weak superstructure reflections at integer and half-integer $l$-positions along for example the (0 5 $l$) line cut (see black arrows in Fig. 4). They appear for even and odd integer $k$ indices and are present in all magnetic phases and also above the magnetic phase transition. Note, the even integer $k$ and half-integer $l$ indices are hidden in Fig. 4(a) underneath the broad Al-powder lines, but are present in the $kl$-plane RSM. This is accordance with the half-integer $l$-indices observed in the RSMs in the $hl$-plane (see Fig. 2). In summary, besides the allowed nuclear Bragg reflections for the space group P*bam*, additional structural Bragg peaks are observed at ($h$ 0 $l$) and (0 $k$ $l$) for odd-integer $h$- and $k$-indices. Further nuclear Bragg peaks are observed at half-integer $l$-positions at (*h, 0, 0.5l*) and (*0, k, 0.5l*) with integer $h$ or $k$ values, respectively. They have not been observed before in the x-ray synchrotron diffraction experiments of Noda *et al.* [32] and Balédent *et al.* [55]. The appearance of the additional nuclear Bragg reflections, which are already present above the magnetic phase transition, indicates that the symmetry of TbMn$_2$O$_5$ must be lower that the previously determined space group P*bam*.

In order to identify the crystal symmetry of TbMn$_2$O$_5$, we have performed an analysis of the extinction rules (reflection conditions) of the possible polar space groups and then modelled the integrated intensities for comparable space groups by performing least-square refinements of the observed nuclear Bragg peaks. As the space group P*bam* can be ruled out, we have considered the *translationsgleiche* subgroups of P*bam*. Among them, P*2$_1$/c* (#14) and P*2/m* (#10) are centrosymmetric and P*2$_1$2$_1$2* (#18) is piezoelectric. They are therefore not applicable. The remaining groups P*c2a* (#32) and P*b2$_1$m* (#26) are polar and allow an electric polarization along *b*. However, the selection rules of our observed nuclear reflections violate three of four extinction rules for P*c2a*. Furthermore, the appearance of odd integer $k$ and integer $l$ indices for (0 $k$ $l$) violates the $k=2n$ extinction rule of the space group P*b2$_1$m*, which was previously suggested to be the symmetry-broken phase (see Table 1 in the Appendix). Therefore, the sub-subgroups have to be considered. Possible polar groups are P*m* (#6), P*2* (#3), P*c* (#7), P*2$_1$* (#4), and P*1* (#1). The observed nuclear Bragg reflections again violate the (0 $k$ $l$), $k=2n$ extinction rule for the space group P*c*. In order to determine the correct space group from the remaining four, the selection rule for the observed (0 $k$ 0), $k=2n$ nuclear Bragg peaks can be utilized. This additional condition is unambiguously fulfilled by P*2$_1$* (P*12$_1$1*) and is ruling out the remaining three space groups (see table 1 in the Appendix). The polar space group P$12_11$ allows an electric polarization along the *b*-direction and enables a doubling of the unit cell along the *c*-direction, which explains the superstructure nuclear Bragg reflections at (*h*, 0, *0.5l*) and (*0, k, 0.5l*) (see the irreducible representations (IRREPs) of P$12_11$ in Table 2 in the Appendix).

To further demonstrate the validity of the assignment to the space group P$12_11$, we have performed a refinement of the integrated intensities of the nuclear Bragg peaks within the commensurate phase measured at 28 K. The integrated intensities of all nuclear Bragg peaks were determined from the full measured RSMs of the *hl*-, *hk*, and *kl*-planes (note, Figs. 2 and 3 only show selected regions). For a comparison with the determined polar space group P$12_11$, we have chosen the centro-symmetric P*bam* and the polar P$b2_1m$ space groups, which were suggested in the literature before. Only the symmetry allowed reflections were considered for each space group. The refinements were performed by least square fits using the FULLPROF program suite [59]. Figure 5 shows the comparison between the calculated and observed intensities for the three considered crystal structures. For the space group P*bam* a residual value of R(F) = 15.9 was obtained. Considerable improvement of the R(F) value was obtained for the polar group P$b2_1m$ with R(F) = 11.2. However, as pointed out above, this space group does not allow the observed (0 *k l*) nuclear Bragg reflections with odd *k* and half integer *l*-indices. For the polar group P$12_11$ an R(F) a slightly higher value of 12.6 was obtained. However, this space group gave the most realistic description of the crystal symmetry of TbMn$_2$O$_5$ with respect to the selection rules. For this space group the superstructure nuclear Bragg reflections along the *c*-direction were also taken into account through displacement modulations. For simplicity the superspace formalism was avoided and instead the irreducible representation of P$12_11$ for the modulation vector k$_N$ = (0, 0, ½) (little group) was used to determine the basis-functions corresponding to the polar vector type distortions (see Table T2 in the Supplemental Material [57]). The best refinement of the structural modulation-distortions was achieved for the IRREP $\Gamma_2$. In this case the structural modulation allows both, Mn$^{3+}$ and Mn$^{4+}$ displacements and also O$^{2-}$ distortions along the *c*-direction and Tb$^{3+}$ displacements along the *a*-direction.

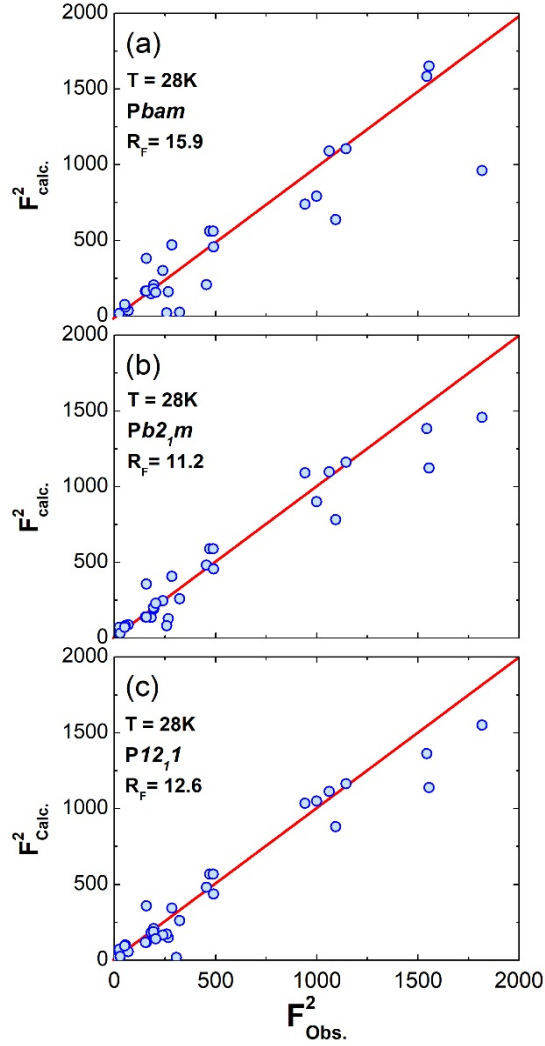

FIG. 5: Experimental and calculated integrated intensities of the fundamental nuclear Bragg reflections. The least square fitting method was applied to model the data taken at 28 K, i.e. in the CM phase of TbMn$_2$O$_5$. The results for the parent space group P*bam* are shown in (a), for the polar P*b*2$_1$*m* group in (b) and the designated polar space group P12$_1$1 in (c).

The selection rules and the results of the refinements shown in Fig. 5 confirm that the polar space group P*12$_1$1* provides the most convincing explanation of our diffraction data. This space group is in contrast to the formerly suggested polar space group P*b*2$_1$*m*, which does not allow (0 *k l*) reflections with odd *k* and half integer *l*-indices and is also not in accordance with the space group P*m* suggested by Balédent *et al.* from single crystal synchrotron diffraction, which however allows all reflections with integer *h*-, *k*-, or *l*-indices [55]. We suggest that the half-integer reflections appear in our neutron diffraction data and not in previous x-ray studies because of the different element-specific coherent scattering cross-sections. For x-ray diffraction using an incident energy of 28 keV, the coherent cross-section for O atoms is quite small. It is $\approx$ 70 times smaller than for Mn and $\approx$ 450 times smaller than Tb. However, the coherent neutron scattering cross sections for these elements are comparable. It is conceivable that the non-integer reflections in our neutron diffraction data arise from coherent scattering from oxygen ions with similar but inequivalent positions. Such displacements could induce electrical dipole moments on the molecular level. However, as in the case of the *a*-direction, they cancel out along the *c*-direction and only contributions along the *b*-axis add up to the final electric polarization.

Below the magnetic phase transition at $T_{HT-ICM}$ = 42 K additional Bragg reflections appear which are not associated with nuclear reflections of the P$12_11$ space group. They can be attributed to magnetic diffraction and are described by the propagation vector of the type $k_M=(1/2+\delta_x, 0, 1/4+\delta_z)$ (see red right angles in Figs. 2 and 3). We have determined the propagation vectors directly from the diffraction maps obtained from the *hl*-plane. At 40 K, magnetic "satellite" Bragg reflections appear at about (-1.51, 0, 2.28) and (-1.51, 0, 2.72) corresponding to $k_M$ and $-k_M$ of the star of the propagation vector for the P$12_11$ space group (see Table T1 in the Supplemental Material [57]). It should be noted that the centrosymmetric space group P*bam* used in previous publications, provides four arms for the star of k which again can be reduced to two if $k_x$ is 0.5 (see Table T2 in the Supplemental Material [57]). With respect to the magnetic selection rules, the magnetic Bragg peaks are in accordance with the crystallographic space group P$12_11$ (see text below Table T2 in the Supplemental Material [57]). Furthermore, the intensity of the magnetic reflections corresponding to each arm of the star of *k* may constitute of the different projections of the magnetic moment perpendicular to the propagation vector [60]. Upon further cooling to the CM phase below $T_{CM}$ = 38 K magnetic "satellite" Bragg reflections shift to (-1.50, 0, 2.25) and (-1.50, 0, 2.75), which corresponds to the commensurate propagation vector $k_M$ = (½, 0, ¼).

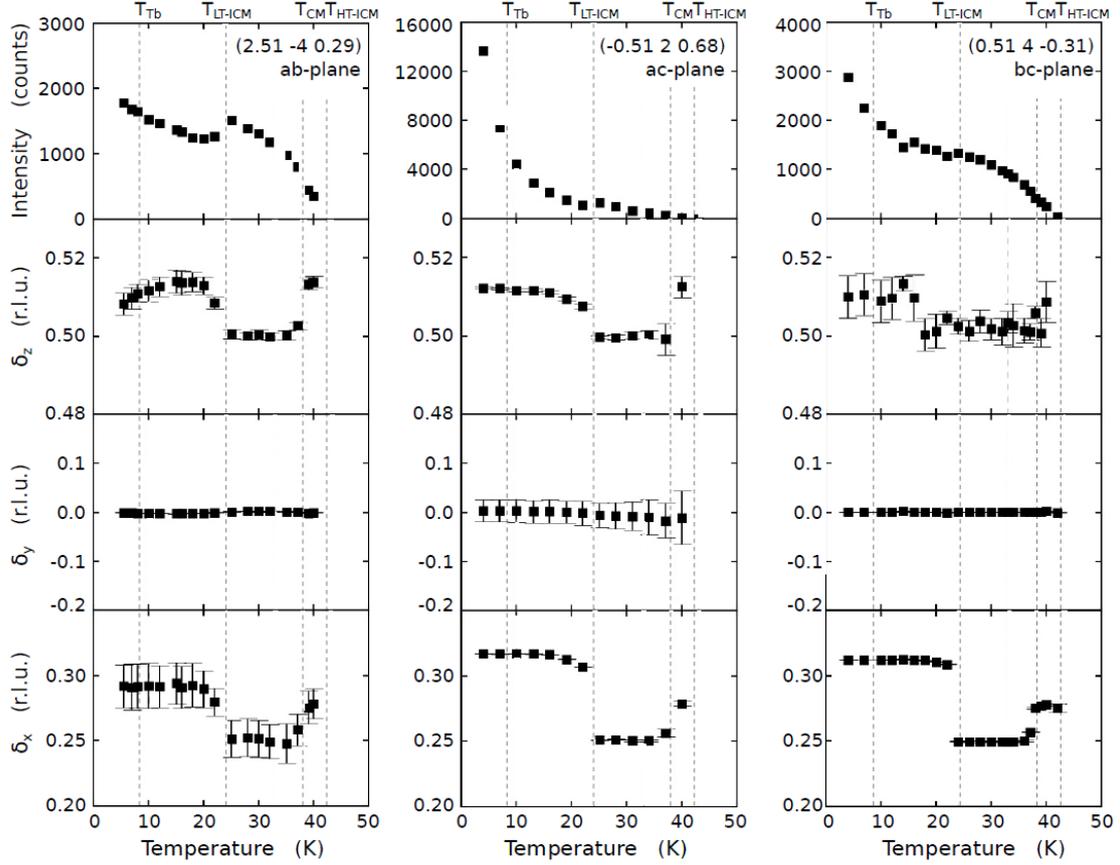

FIG. 6. Temperature dependence of the $q_h$, $q_k$, and $q_l$ peak positions and peak intensities of selected magnetic Bragg reflections determined from the neutron diffraction measurements with the crystal being aligned in the *ab*-, *ac*-, and *bc*-planes. Note that the $q_l$ position for the measurement in the *ab*-plane, the $q_k$ position in the *ac*-plane, $q_h$ in the *bc*-plane were difficult to be determined precisely since they appear out of the scattering plane. The different magnetic phase transitions are indicated by vertical dashed lines.

The temperature dependence of the peak positions in the $q_h$-, $q_k$-, and $q_l$-directions and of the intensity of selected magnetic Bragg peaks, i.e. of (2.51, -4, 0.29), (-0.51, 2, 0.69), and (0.51, 4, -0.31) are shown in Fig. 6. The data were obtained by fitting Gaussian profiles to the peaks obtained from line cuts through the RSMs of the diffraction measurements in the *ab*-, *ac*-, and *bc*-planes, respectively. The magnetic phase transitions occur at $T_{N1}$ = 42(1) K to the HT-ICM phase, at $T_{CM}$ = 38(1) K to the commensurate phase, and $T_{LT-ICM}$ = 24(1) K to the low temperature incommensurate phase and are indicated by vertical dashed lines. The transition temperatures are in good agreement with previously published neutron diffraction data of Kobayashi *et al.* [28], Chapon *et al.* [22], and Wilkinson *et al.* [29]. However, Kobayashi *et al.* [28] observed a coexistence of the HT-ICM and CM phase from 37.8 K to 36.5 K and of the CM and LT-ICM phase from 24.0 to 22.4 K. A similar result was obtained by Wilkinson *et al.* [29]. This was not seen in our experiment probably due to the coarser temperature steps used. At the onset of the magnetic order at $T_{HT-ICM}$ = 42(1) K magnetic Bragg peaks appear at ($h\pm0.518$, $k$, $l\pm0.28$) with $h$ being an even integer and $l$ an integer number. In the commensurate phase between 38(1) K and 24(1) K the magnetic propagation vector locks in at ($h\pm0.5$, $k$, $l\pm0.25$) and abruptly increases to ($h\pm0.508$, $k$, $l\pm0.309$) at the onset of the LT-ICM phase below $T_{LT-ICM}$ = 24(1) K. It increases upon further cooling and saturates at ($h\pm0.513$, $k$, $l\pm0.313$) below about 16 K. This is in good agreement with previously published data [22,28,29]. Our data do not provide any evidence for an incommensurability of the magnetic Bragg reflections in the CM phase. Remarkably, the intensities of the different magnetic Bragg peaks shown in Fig. 6 possess a distinct behavior. From 42 K to 24 K, i.e. in the HT-ICM and in the CM phase the intensities of all three magnetic Bragg peaks increase simultaneously. In this temperature range predominantly the Mn-moments are ordered. At the transition to the LT-ICM phase at 24 K the intensities of the (2.51, -4, 0.29) and (-0.51, 2, 0.69) first decrease and then increase steeply upon further cooling while the (0.51, 4, -0.31) Bragg reflection increases continuously. Furthermore, the (-0.51, 2, 0.69) Bragg peak shows the strongest increase in intensity at lowest temperature. Compared with [29], it is most likely that the magnetic moments of $Mn^{3+}/Mn^{4+}$ gain a *c*-component at $T_{LT-ICM}$ reducing the magnetic component, specially of $Mn^{3+}$ on the *ab*-plane and a *bc* cycloid is realized within a more complicated magnetic structure. Already about 10 K above the expected ordering temperature of the $Tb^{3+}$ moments the magnetic exchange between the Mn- and Tb-moments leads to a polarization of the Tb-spins. This has also been observed before in related materials such as $DyMnO_3$ [40], $TbVO_3$ [61], or $CeVO_3$ [62]. Wilkinson *et al.* determined that there are two distinct Tb-sublattices in $TbMn_2O_5$, which possess a different increase in magnetic moment and a different rotation of the Tb-spins in the *ab*-plane with temperature [29]. This can explain the observed differences in the temperature dependencies of the intensity of the different magnetic Bragg reflections, in particular the rapid increase at lowest temperatures due to the contribution again from the *ab* plane but from $Tb^{3+}$ moments. The increase in the peak intensities from 20 K to 4 K is similar in form to the measured electrical polarization [7,8], indicating that both phenomena are related to each other. The existence of a spin-spiral in all magnetic phases of $TbMn_2O_5$ with a propagation direction along the *c*-axis and close to quadrature phase steps has already been confirmed by previous diffraction experiments [29,32,37]. Due to the corresponding spin-current the induced electrical polarization is proportional to $S_i \times S_j$. This demonstrates that the leading mechanism for the magnetoelectric coupling in the LT-ICM phase is the spin-current while, as discussed above, the dominating mechanism for the ME coupling in the CM phase are atomic displacements caused by exchange-striction proportional to $S_i \cdot S_j$.

## IV. CONCLUSIONS

In summary, our single crystal neutron diffraction experiments provide direct evidence of a lowering of the crystallographic symmetry for TbMn$_2$O$_5$ from the previously determined crystal structure P*bam* to P$12_1$1 through the observation of additional structural Bragg peaks at ($h$ 0 $l$) and (0 $k$ $l$) with $h$ or $k$ being an odd integer and $l$ an integer number and at ($h$ 0 $l$) and (0 $k$ $l$) positions with $h$ or $k$ being an integer and $l$ a half integer index. This structure exists already above the magnetic ordering temperature and does not change through all magnetic phases down to the lowest temperature. A non-centrosymmetric space group describing the precise crystallographic structure gives validity to the $S_i \cdot S_j$ based exchange-striction description of magnetoelectric coupling within the commensurate magnetic phase. In this scenario either the Mn$^{3+}$ or Mn$^{4+}$ ions move in direction of the *ab*-plane zigzag units. This lifts the degeneracy of the magnetic exchange interaction J$_3$, which links the zigzag chains and hence lowers the magnetic frustration and therefore stabilizes the magnetic structure. The atomic displacements generate electric polarizations on the atomic level, which cancel out in *a*-direction and accumulate along the *b*-direction. Previously unobserved nuclear reflections at half-positions in P*bam* suggest a superstructure with a doubling of the unit cell exists along the *c*-direction. The observation of these Bragg peaks only in neutron diffraction measurements suggests that these reflections correspond to a subtle displacement between otherwise inequivalent oxygen positions. This can create an electric polarization at the atomic level, which, however cancel out along the *c*-direction. In the LT-ICM phase the intensity of the magnetic Bragg reflections follows the intensity of the electric polarization. This suggests that the leading mechanism for the ME-coupling in the LT-ICM phases is the spin-current model, which is caused by a spin-cycloid in the *bc*-plane with a propagation direction along the c-axis. In this case the electric polarization is proportional to $S_i \times S_j$. These observations bring clarity to the previous issues regarding magnetoelectric coupling in TbMn$_2$O$_5$, which can also be readily generalized to the entire multiferroic RMn$_2$O$_5$ series.

**ACKNOWLEDGMENTS**

This work was supported through the Australian Institute of Nuclear Science and Engineering Ltd (AINSE) and the Australian Research Council (ARC) through the funding of the Discovery Grants DP110105346 and DP160100545. The support from the Deutsche Forschungsgemeinschaft through project B01 of the Collaborative Research Center 1143 (Project No. 247310070) and the Würzburg-Dresden Cluster of Excellence on Complexity and Topology in Quantum Matter-ct.qmat (Project No. 390858490) is gratefully acknowledged. We thank the Australian Centre for Neutron Scattering for the allocation of neutron beam time on the instruments Joey and Wombat.

**APPENDIX: SYMMETRY ANAYSIS OF THE DIFFERENT SPACE GROUPS**

TABLE 1: Extinction rules for the space groups of interest. The reflection conditions are listed in column 3 and 4 (from Ref. [16]).

| Space group | Point group | General | Special |
|---|---|---|---|
| P*bam* | mmm | 0kl, k=2n<br>h0l, h=2n<br>h00, h=2n<br>0k0, k=2n | hkl, h+k=2n (for 2a, 2b, 2c, 2d, 4e, 4f) |
| P*b*2$_1$*m* | m2m | 0kl, k=2n<br>0k0, k=2n | |
| P*c*2*a* | m2m | hk0, h=2n<br>0kl, l=2n<br>00l, l=2n<br>h00, h=2n | |
| P12$_1$1 (unique axis *b*) | 2 | 0k0, k=2n | |
| P*c* | m | 0kl, k=2n<br>0k0, k=2n | |
| P*m* | m | no | |
| P*2* | 2 | no | |
| P*1* | 1 | no | |

TABLE 2: Irreducible representations (IRREPs) of $P12_11$ for $k_N=(0, 0, 1/2)$ and the basis for polar vector type distortions (for all atoms on the Wyckoff position $2a$). The little group contains both of the symmetry elements belonging to $P12_11$ (from Ref. [16]).

| IRREPs | Characters | | Basis functions | |
| --- | --- | --- | --- | --- |
| | 1 | $2_1$ | x,y,z | -x+1,y+1/2,-z+1 |
| $\Gamma_1$ | 1 | 1 | (u,v,w) | (u,-v,w) |
| $\Gamma_2$ | 1 | -1 | (u,v,w) | (-u,v,-w) |

Decomposition of the representation: $\Gamma_{Dis,2a}=3\cdot\Gamma_1+3\cdot\Gamma_2$

# Reduced crystal symmetry as origin of the ferroelectric polarization within the incommensurate magnetic phase of TbMn$_2$O$_5$

N. Narayanan[1,2], P. J. Graham[3], P. Rovillain[2,3,4], J. O'Brien[3], J. Bertinshaw[3,5], S. Yick[2,3], J. Hester[2], A. Maljuk[6,7], D. Souptel[7], B. Büchner[7,8,9], D. Argyriou[6,10], and C. Ulrich[3,*]

[1]*School of Physical, Environmental and Mathematical Sciences, University of New South Wales, Canberra, Australian Capital Territory 2600, Australia*

[2]*The Australian Centre for Neutron Scattering, Australian Nuclear Science and Technology Organisation, Lucas Heights, New South Wales 2234, Australia*

[3]*School of Physics, University of New South Wales, Sydney, NSW 2052, Australia*

[4]*Sorbonne Universitée, CNRS, Institut des NanoSciences de Paris, INSP, UMR7588, F-75005 Paris, France.*

[5]*Max-Planck-Institut für Festkörperforschung, Heisenbergstrasse 1, D-70569 Stuttgart, Germany*

[6]*Helmholtz-Zentrum Berlin für Materialien und Energie, D-14109 Berlin, Germany*

[7]*Leibniz Institut für Festkörper- und Werkstoffforshung (IFW) Dresden, Helmholtzstr. 20, D-01069 Dresden, Germany*

[8]*Institut für Festkörper- und Materialphysik, Technische Universität Dresden, 01069 Dresden, Germany*

[9]*Würzburg-Dresden Cluster of Excellence ct.qmat, Technische Universität Dresden, 01062 Dresden, Germany*

[10]*European Spallation Source ESS AB, S-22100 Lund, Sweden*

(Dated: 11 September 2021)

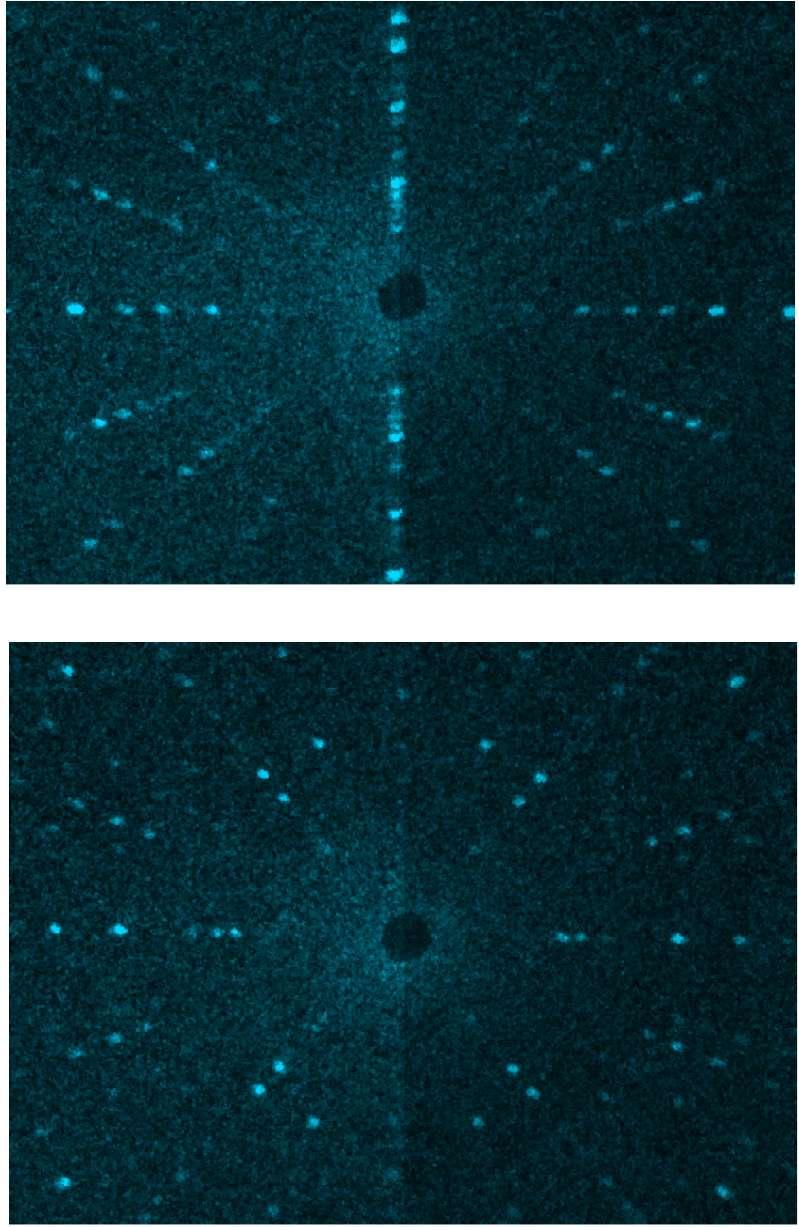

FIG. S1 Neutron Laue diffraction on the TbMn$_2$O$_5$ single crystal measured on the instrument JOEY at the ANSTO. In the upper panel the crystal was oriented with the incident beam along the [001] direction and the lower panel along the [110] directions. The data demonstrate the high quality of the single crystal and the absence of twining within the instrumental resolution.

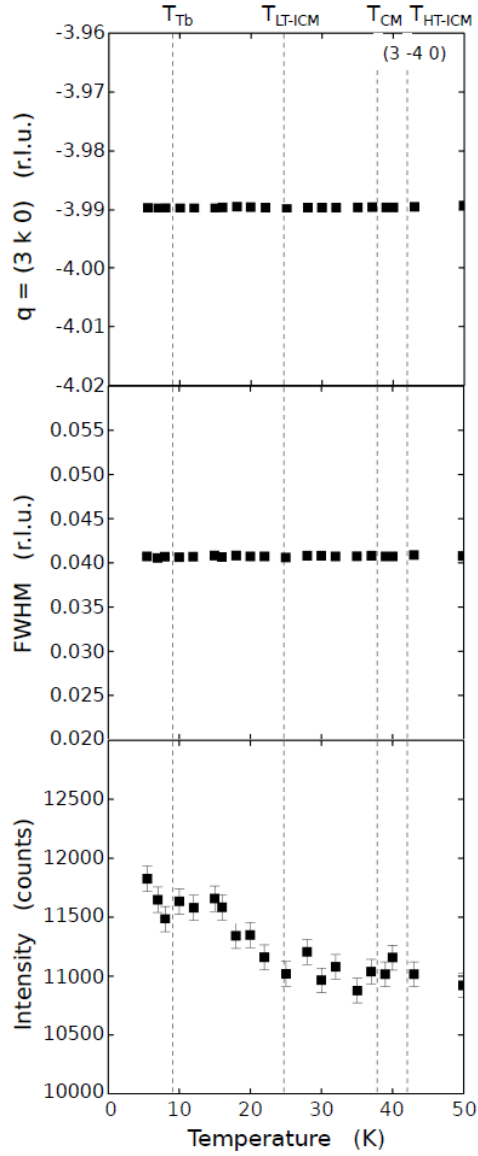

*FIG. S2 Temperature dependence of the position in $q_I$-direction, linewidth (FWHM: full width at half maximum) and peak intensity of the (3 -4 0) Bragg peak. This peak is not allowed in the Pbam structure. It is important to note that the peak is already observed above the magnetic phase transitions and remains unchanged in all magnetic phases, the HT-ICM, CM, and LT-CM phase.*

TABLE T1: The star of magnetic Bragg peaks at $k_M = (k_x, 0, k_z)$ within the space group $P12_11$. Only the rotation part of the symmetry element was used.

| Symmetry element | Initial propagation vector | Transformed propagation vector |
|---|---|---|
| 1 | $\begin{pmatrix} k_x \\ 0 \\ k_z \end{pmatrix}$ | $\begin{pmatrix} k_x \\ 0 \\ k_z \end{pmatrix}$ |
| $2_{1,y}$ | $\begin{pmatrix} k_x \\ 0 \\ k_z \end{pmatrix}$ | $\begin{pmatrix} -k_x \\ 0 \\ -k_z \end{pmatrix}$ |

TABLE T2: The star of magnetic Bragg peaks $k_M = (k_x, 0, k_z)$ within the space group P*bam*. Only the rotation part of the symmetry element was used.

| Symmetry element | Initial propagation vector | Transformed propagation vector |
|---|---|---|
| 1 | $\begin{pmatrix} k_x \\ 0 \\ k_z \end{pmatrix}$ | $\begin{pmatrix} k_x \\ 0 \\ k_z \end{pmatrix}$ |
| $\bar{1}$ | $\begin{pmatrix} k_x \\ 0 \\ k_z \end{pmatrix}$ | $\begin{pmatrix} -k_x \\ 0 \\ -k_z \end{pmatrix}$ |
| $2_{1,x}$ | $\begin{pmatrix} k_x \\ 0 \\ k_z \end{pmatrix}$ | $\begin{pmatrix} k_x \\ 0 \\ -k_z \end{pmatrix}$ |
| $2_{1,y}$ | $\begin{pmatrix} k_x \\ 0 \\ k_z \end{pmatrix}$ | $\begin{pmatrix} -k_x \\ 0 \\ -k_z \end{pmatrix}$ |
| $2_{1,z}$ | $\begin{pmatrix} k_x \\ 0 \\ k_z \end{pmatrix}$ | $\begin{pmatrix} -k_x \\ 0 \\ k_z \end{pmatrix}$ |
| m(xy) | $\begin{pmatrix} k_x \\ 0 \\ k_z \end{pmatrix}$ | $\begin{pmatrix} k_x \\ 0 \\ -k_z \end{pmatrix}$ |
| a(x ¼ z) | $\begin{pmatrix} k_x \\ 0 \\ k_z \end{pmatrix}$ | $\begin{pmatrix} k_x \\ 0 \\ k_z \end{pmatrix}$ |
| b(¼ y z) | $\begin{pmatrix} k_x \\ 0 \\ k_z \end{pmatrix}$ | $\begin{pmatrix} -k_x \\ 0 \\ k_z \end{pmatrix}$ |

For both commensurate and incommensurate propagation vectors of the type (1/2, 0, 1/4) and (1/2+$\delta_x$, 0, 1/4+$\delta_z$), within *P12_11* the little group consists of the identity operator. The difference between the cases is shown in the equivalence and non-equivalence of $k_M$ and -$k_M$ respectively. There is only one one-dimensional magnetic IRREP $\Gamma_{1,\,Mag}$ (character $\chi=1$) with basis vectors consisting of all three components.

Decomposition of the magnetic IRREP: $\Gamma_{Mag,\,2a}=3\cdot\Gamma_{1,\,Mag}$

Basis vectors: (1 0 0), (0 1 0), (0 0 1)

Each of the two Wyckoff sites of the ions $Mn^{3+}$, $Mn^{4+}$ and $Tb^{3+}$ within *P12_11* split further into two positions. The phase between the magnetic moments on each of the pair of positions can be adjusted to be compatible with the experimentally known values and extinction rules within *Pbam* used in literature.